
\magnification=\magstep1
\newbox\SlashedBox
\def\slashed#1{\setbox\SlashedBox=\hbox{#1}
\hbox to 0pt{\hbox to 1\wd\SlashedBox{\hfil/\hfil}\hss}#1}
\def\hboxtosizeof#1#2{\setbox\SlashedBox=\hbox{#1}
\hbox to 1\wd\SlashedBox{#2}}

\def\mathslashed#1{\setbox\SlashedBox=\hbox{$#1$}
\hbox to 0pt{\hbox to 1\wd\SlashedBox{\hfil/\hfil}\hss}#1}

\def\ifsmall{\iffalse}  
\def\titlepagefont{}  

\def\DefineTeXgraphics{%
\special{ps::[global] /TeXgraphics { } def}}  

\def\today{\ifcase\month\or January\or February\or March\or April\or May
\or June\or July\or August\or September\or October\or November\or
December\fi\space\number\day, \number\year}
\def\eatPrefix19{}
\def\Year{\expandafter\eatPrefix\the\year}
\newcount\hours \newcount\minutes
\def\monthname{\ifcase\month\or
January\or February\or March\or April\or May\or June\or July\or
August\or September\or October\or November\or December\fi}
\def\shortmonthname{\ifcase\month\or
Jan\or Feb\or Mar\or Apr\or May\or Jun\or Jul\or
Aug\or Sep\or Oct\or Nov\or Dec\fi}

\def\TimeStamp{\hours\the\time\divide\hours by60%
\minutes -\the\time\divide\minutes by60\multiply\minutes by60%
\advance\minutes by\the\time%
${\rm \shortmonthname}\cdot\the\day\cdot\the\year%
\qquad\the\hours:\the\minutes$}




\def\Title#1{%
\vskip 1in{\titlefont\centerline{#1}}\vskip .5in}


\newdimen\fullhsize
\newdimen\hstitle
\hstitle=\hsize 
\newdimen\hsbody
\hsbody=\hsize 
\newdimen\hbodyoffset
\hbodyoffset=\hoffset 
\newbox\leftpage
\def\abstract#1{#1}
\def\rotated{\special{ps: landscape}
\magnification=1000  
\baselineskip=14pt
\global\hstitle=9truein\global\hsbody=4.75truein
\global\vsize=7truein\global\voffset=-.31truein
\global\hoffset=-0.54in\global\hbodyoffset=-.54truein
\global\fullhsize=10truein
\def\DefineTeXgraphics{%
\special{ps::[global]
/TeXgraphics {currentpoint translate 0.7 0.7 scale
              -80 0.72 mul -1000 0.72 mul translate} def}}
\let\lr=L
\def\ifsmall{\iftrue}
\def\titlepagefont{\twelvepoint}
\trueseventeenpoint
\def\almostshipout##1{\if L\lr \count1=1
      \global\setbox\leftpage=##1 \global\let\lr=R
   \else \count1=2
      \shipout\vbox{\hbox to\fullhsize{\box\leftpage\hfil##1}}
      \global\let\lr=L\fi}

\output={\ifnum\count0=1 
 \shipout\vbox{\hbox to \fullhsize{\hfill\pagebody\hfill}}\advancepageno
 \else
 \almostshipout{\leftline{\vbox{\pagebody\makefootline}}}\advancepageno
 \fi}

\def\abstract##1{{\leftskip=1.5in\rightskip=1.5in ##1\par}} }

\global\newcount\secno \global\secno=0
\global\newcount\appno \global\appno=0
\global\newcount\meqno \global\meqno=1
\global\newcount\figno \global\figno=0

\def\Section#1{\global\advance\secno by1\relax\global\meqno=1
\bigbreak\bigskip
\centerline{\twelvepoint \bf %
\the\secno. #1}%
\par\nobreak\medskip\nobreak}
\def\tagsection#1{\edef#1{\the\secno}%
\ifWritingAuxFile\immediate\write\auxfile{\noexpand\xdef\noexpand#1{#1}}\fi%
}
\def\section{\Section}

\def\romappno{\uppercase\expandafter{\romannumeral\appno}}
\def\Appendix#1{\global\advance\appno by1\relax\global\meqno=1\global\secno=0
\bigbreak\bigskip
\centerline{\twelvepoint \bf Appendix %
A. #1}%
\par\nobreak\medskip\nobreak}
\def\tagappendix#1{\edef#1{\romappno}%
\ifWritingAuxFile\immediate\write\auxfile{\noexpand\xdef\noexpand#1{#1}}\fi%
}
\def\appendix{\Appendix}

\def\eqn#1{
\ifnum\secno>0
  \eqno(\the\secno.\the\meqno)\xdef#1{\the\secno.\the\meqno}%
     \global\advance\meqno by1
\else\ifnum\appno>0
  \eqno({\rm\romappno}.\the\meqno)\xdef#1{\romappno.\the\meqno}%
     \global\advance\meqno by1
\else
  \eqno(\the\meqno)\xdef#1{\the\meqno}%
     \glovbal\advance\meqno by1
\fi\fi%
\ifWritingAuxFile\immediate\write\auxfile{\noexpand\xdef\noexpand#1{#1}}\fi%
}
\def\equn{
\ifnum\secno>0
  \global\advance\meqno by1
  \eqno(\the\meqno)
\else\ifnum\appno>0
  \eqno(A.\the\meqno)%
     \global\advance\meqno by1
\else
  \eqno(\the\meqno)%
     \global\advance\meqno by1
\fi\fi%
}
\def\defeqn#1{
\ifnum\secno>0
  \xdef#1{\the\secno.\the\meqno}%
     \global\advance\meqno by1
\else\ifnum\appno>0
  \xdef#1{\romappno.\the\meqno}%
     \global\advance\meqno by1
\else
  \xdef#1{\the\meqno}%
     \global\advance\meqno by1
\fi\fi%
\ifWritingAuxFile\immediate\write\auxfile{\noexpand\xdef\noexpand#1{#1}}\fi%
}
\def\anoneqn{
\ifnum\secno>0
  \eqno(\the\secno.\the\meqno)%
     \global\advance\meqno by1
\else\ifnum\appno>0
  \eqno({\rm\romappno}.\the\meqno)%
     \global\advance\meqno by1
\else
  \eqno(\the\meqno)%
     \global\advance\meqno by1
\fi\fi%
}
\def\mfig#1#2{\global\advance\figno by1%
\relax#1\the\figno\edef#2{\the\figno}%
\ifWritingAuxFile\immediate\write\auxfile{\noexpand\xdef\noexpand#2{#2}}\fi%
}

\catcode`@=11 

\font\ninerm=cmr9
\font\eightrm=cmr8
\font\sixrm=cmr6

\def\loadtrueseventeenpoint{
 \font\seventeenrm=cmr10 at 17.28truept
 \font\seventeeni=cmmi10 at 17.28truept
 \font\seventeenbf=cmbx10 at 17.28truept
 \font\seventeenit=cmti10 at 17.28truept
 \font\seventeensl=cmsl10 at 17.28truept
 \font\seventeensy=cmsy10 at 17.28truept
}
\def\loadfourteenpoint{
\font\fourteenrm=cmr10 at 14.4pt
\font\fourteeni=cmmi10 at 14.4pt
\font\fourteenit=cmti10 at 14.4pt
\font\fourteensl=cmsl10 at 14.4pt
\font\fourteensy=cmsy10 at 14.4pt
\font\fourteenbf=cmbx10 at 14.4pt
}
\def\loadtruetwelvepoint{
\font\twelverm=cmr10 at 12truept
\font\twelvei=cmmi10 at 12truept
\font\twelveit=cmti10 at 12truept
\font\twelvesl=cmsl10 at 12truept
\font\twelvesy=cmsy10 at 12truept
\font\twelvebf=cmbx10 at 12truept
}

\font\ninei=cmmi9
\font\eighti=cmmi8
\font\sixi=cmmi6
\skewchar\ninei='177 \skewchar\eighti='177 \skewchar\sixi='177

\font\ninesy=cmsy9
\font\eightsy=cmsy8
\font\sixsy=cmsy6
\skewchar\ninesy='60 \skewchar\eightsy='60 \skewchar\sixsy='60

\font\ninebf=cmbx9
\font\eightbf=cmbx8
\font\sixbf=cmbx6

\font\ninett=cmtt9
\font\eighttt=cmtt8

\hyphenchar\tentt=-1 
\hyphenchar\ninett=-1
\hyphenchar\eighttt=-1

\font\ninesl=cmsl9
\font\eightsl=cmsl8

\font\nineit=cmti9
\font\eightit=cmti8


\newskip\ttglue
\def\tenpoint{\def\rm{\fam0\tenrm}%
  \textfont0=\tenrm \scriptfont0=\sevenrm \scriptscriptfont0=\fiverm
  \textfont1=\teni \scriptfont1=\seveni \scriptscriptfont1=\fivei
  \textfont2=\tensy \scriptfont2=\sevensy \scriptscriptfont2=\fivesy
  \textfont3=\tenex \scriptfont3=\tenex \scriptscriptfont3=\tenex
  \def\it{\fam\itfam\tenit}\textfont\itfam=\tenit
  \def\sl{\fam\slfam\tensl}\textfont\slfam=\tensl
  \def\bf{\fam\bffam\tenbf}\textfont\bffam=\tenbf \scriptfont\bffam=\sevenbf
  \scriptscriptfont\bffam=\fivebf
  \normalbaselineskip=12pt
  \let\sc=\eightrm
  \let\big=\tenbig
  \setbox\strutbox=\hbox{\vrule height8.5pt depth3.5pt width\z@}%
  \normalbaselines\rm}

\def\twelvepoint{\def\rm{\fam0\twelverm}%
  \textfont0=\twelverm \scriptfont0=\ninerm \scriptscriptfont0=\sevenrm
  \textfont1=\twelvei \scriptfont1=\ninei \scriptscriptfont1=\seveni
  \textfont2=\twelvesy \scriptfont2=\ninesy \scriptscriptfont2=\sevensy
  \textfont3=\tenex \scriptfont3=\tenex \scriptscriptfont3=\tenex
  \def\it{\fam\itfam\twelveit}\textfont\itfam=\twelveit
  \def\sl{\fam\slfam\twelvesl}\textfont\slfam=\twelvesl
  \def\bf{\fam\bffam\twelvebf}\textfont\bffam=\twelvebf
                                           \scriptfont\bffam=\ninebf
  \scriptscriptfont\bffam=\sevenbf
  \normalbaselineskip=12pt
  \let\sc=\eightrm
  \let\big=\tenbig
  \setbox\strutbox=\hbox{\vrule height8.5pt depth3.5pt width\z@}%
  \normalbaselines\rm}

\def\fourteenpoint{\def\rm{\fam0\fourteenrm}%
  \textfont0=\fourteenrm \scriptfont0=\tenrm \scriptscriptfont0=\sevenrm
  \textfont1=\fourteeni \scriptfont1=\teni \scriptscriptfont1=\seveni
  \textfont2=\fourteensy \scriptfont2=\tensy \scriptscriptfont2=\sevensy
  \textfont3=\tenex \scriptfont3=\tenex \scriptscriptfont3=\tenex
  \def\it{\fam\itfam\fourteenit}\textfont\itfam=\fourteenit
  \def\sl{\fam\slfam\fourteensl}\textfont\slfam=\fourteensl
  \def\bf{\fam\bffam\fourteenbf}\textfont\bffam=\fourteenbf%
  \scriptfont\bffam=\tenbf
  \scriptscriptfont\bffam=\sevenbf
  \normalbaselineskip=17pt
  \let\sc=\elevenrm
  \let\big=\tenbig
  \setbox\strutbox=\hbox{\vrule height8.5pt depth3.5pt width\z@}%
  \normalbaselines\rm}

\def\seventeenpoint{\def\rm{\fam0\seventeenrm}%
  \textfont0=\seventeenrm \scriptfont0=\fourteenrm \scriptscriptfont0=\tenrm
  \textfont1=\seventeeni \scriptfont1=\fourteeni \scriptscriptfont1=\teni
  \textfont2=\seventeensy \scriptfont2=\fourteensy \scriptscriptfont2=\tensy
  \textfont3=\tenex \scriptfont3=\tenex \scriptscriptfont3=\tenex
  \def\it{\fam\itfam\seventeenit}\textfont\itfam=\seventeenit
  \def\sl{\fam\slfam\seventeensl}\textfont\slfam=\seventeensl
  \def\bf{\fam\bffam\seventeenbf}\textfont\bffam=\seventeenbf%
  \scriptfont\bffam=\fourteenbf
  \scriptscriptfont\bffam=\twelvebf
  \normalbaselineskip=21pt
  \let\sc=\fourteenrm
  \let\big=\tenbig
  \setbox\strutbox=\hbox{\vrule height 12pt depth 6pt width\z@}%
  \normalbaselines\rm}

\def\ninepoint{\def\rm{\fam0\ninerm}%
  \textfont0=\ninerm \scriptfont0=\sixrm \scriptscriptfont0=\fiverm
  \textfont1=\ninei \scriptfont1=\sixi \scriptscriptfont1=\fivei
  \textfont2=\ninesy \scriptfont2=\sixsy \scriptscriptfont2=\fivesy
  \textfont3=\tenex \scriptfont3=\tenex \scriptscriptfont3=\tenex
  \def\it{\fam\itfam\nineit}\textfont\itfam=\nineit
  \def\sl{\fam\slfam\ninesl}\textfont\slfam=\ninesl
  \def\bf{\fam\bffam\ninebf}\textfont\bffam=\ninebf \scriptfont\bffam=\sixbf
  \scriptscriptfont\bffam=\fivebf
  \normalbaselineskip=11pt
  \let\sc=\sevenrm
  \let\big=\ninebig
  \setbox\strutbox=\hbox{\vrule height8pt depth3pt width\z@}%
  \normalbaselines\rm}

\def\eightpoint{\def\rm{\fam0\eightrm}%
  \textfont0=\eightrm \scriptfont0=\sixrm \scriptscriptfont0=\fiverm%
  \textfont1=\eighti \scriptfont1=\sixi \scriptscriptfont1=\fivei%
  \textfont2=\eightsy \scriptfont2=\sixsy \scriptscriptfont2=\fivesy%
  \textfont3=\tenex \scriptfont3=\tenex \scriptscriptfont3=\tenex%
  \def\it{\fam\itfam\eightit}\textfont\itfam=\eightit%
  \def\sl{\fam\slfam\eightsl}\textfont\slfam=\eightsl%
  \def\bf{\fam\bffam\eightbf}\textfont\bffam=\eightbf \scriptfont\bffam=\sixbf%
  \scriptscriptfont\bffam=\fivebf%
  \normalbaselineskip=9pt%
  \let\sc=\sixrm%
  \let\big=\eightbig%
  \setbox\strutbox=\hbox{\vrule height7pt depth2pt width\z@}%
  \normalbaselines\rm}

\def\tenbig#1{{\hbox{$\left#1\vbox to8.5pt{}\right.\n@space$}}}
\def\ninebig#1{{\hbox{$\textfont0=\tenrm\textfont2=\tensy
  \left#1\vbox to7.25pt{}\right.\n@space$}}}
\def\eightbig#1{{\hbox{$\textfont0=\ninerm\textfont2=\ninesy
  \left#1\vbox to6.5pt{}\right.\n@space$}}}

\def\footnote#1{\edef\@sf{\spacefactor\the\spacefactor}#1\@sf
      \insert\footins\bgroup\eightpoint
      \interlinepenalty100 \let\par=\endgraf
        \leftskip=\z@skip \rightskip=\z@skip
        \splittopskip=10pt plus 1pt minus 1pt \floatingpenalty=20000
        \smallskip\item{#1}\bgroup\strut\aftergroup\@foot\let\next}
\skip\footins=12pt plus 2pt minus 4pt 
\dimen\footins=30pc 

\newinsert\margin
\dimen\margin=\maxdimen
\def\titlefont{\seventeenpoint}
\loadtruetwelvepoint 
\loadtrueseventeenpoint
\catcode`\@=\active
\catcode`\"=\active

\def\eatOne#1{}
\def\ifundef#1{\expandafter\ifx%
\csname\expandafter\eatOne\string#1\endcsname\relax}


\global\newcount\refno \global\refno=1
\newwrite\rfile
\newlinechar=`\^^J
\def\ref#1#2{\the\refno\nref#1{#2}}
\def\nref#1#2{\xdef#1{\the\refno}%
\ifnum\refno=1\immediate\openout\rfile=refs.tmp\fi%
\immediate\write\rfile{\noexpand\item{[\noexpand#1]\ }#2.}%
\global\advance\refno by1}
\def\lref#1#2{\the\refno\xdef#1{\the\refno}%
\ifnum\refno=1\immediate\openout\rfile=refs.tmp\fi%
\immediate\write\rfile{\noexpand\item{[\noexpand#1]\ }#2\semi}%
\global\advance\refno by1}
\def\cref#1{\immediate\write\rfile{#1\semi}}

\def\semi{;\hfil\noexpand\break}

\newif\ifWritingAuxFile
\newwrite\auxfile
\def\SetUpAuxFile{%
\xdef\auxfileName{\jobname.aux}%
\openin1 \auxfileName \ifeof1\message{No file \auxfileName; I'll create one.
}\else\closein1\relax\input\auxfileName\fi%
\WritingAuxFiletrue%
\immediate\openout\auxfile=\auxfileName}

\def\L{\left(}

\def\tr{\hbox{\rm tr}}
\def\mod{\hbox{\rm mod}}

\def\La{\Lambda}
\def\la{\lambda}

\def\quotesr{''}

\def\ref#1#2{\nref#1{#2}}
\overfullrule 0pt
\hfuzz 52pt
\hsize 6.50 truein
\vsize 8.5 truein

\loadfourteenpoint
\newcount\meqno
\newcount\secno
\meqno=0
\secno=0
\def\secta{\global\advance\secno by1
}

\def\put#1{\global\edef#1{(\the\meqno)}     }

\def\cgko{1}
\def\cbpz{2}
\def\cfms{3}
\def\cmoore{4}
\def\cgep{5}
\def\cshellb{6}
\def\ckac{7}
\def\cgod{8}
\def\cverlinde{9}
\def\cshella{10}
\def\cintr{11}
\def\cbais{12}
\def\cextended{13}
\def\ccardya{14}
\def\cshellc{15}
\def\cprevious{16}
\def\ccappelli{17}
\def\czam{18}
\def\cpatera{19}


\def\Tabethree{1}
\def\Tabeseven{2}
\def\Tabesix{3}
\def\Tabefour{4}
\def\Tabefive{5}

\def\L{\Lambda}

\def\asu{\widehat{su}}
\def\aso{\widehat{so}}

\def\suson{\asu(N)_2/\aso(N)_4}
\def\dser{\aso(2N)_2/\aso(N)_2\times\aso(N)_2}
\noindent   \hfill LTH-93-310 \break

\noindent   \hfill SWAT-93-07 \break

\noindent \hfill   June 93 \break

\vskip - 1 cm

\Title{\vbox{\centerline{ Maverick Examples Of }
\centerline{ Coset Conformal Field Theories }}}

\vskip .2 cm
\centerline{\bf David C. Dunbar }

\centerline{\it Dept. of Physics}
\centerline{\it University College of Swansea}
\centerline{\it Swansea, Wales, SA2 8PP, U.K.}

\vskip 0.5 truecm

\centerline{\bf and }

\vskip 0.5 truecm

\centerline{ \bf  Keith G. Joshi }

\centerline{\it D.A.M.T.P.}
\centerline{\it University of Liverpool }
\centerline{\it Liverpool, L69 3BX, U.K.}

\vskip 1.0 cm

\vskip 2.0 truecm \baselineskip12pt

\centerline{\bf Abstract }

\vskip 0.3 truecm
{
We present coset conformal field theories
whose spectrum is not determined by the identification
current method.
In these ``maverick'' cosets there is
a larger symmetry
identifying primary fields
than under the identification current.
We find an A-D-E classification of these mavericks. }
\narrower\smallskip

\baselineskip14pt

\vfill

\break




\vskip .4 truecm
\secta
\noindent
{\bf  Introduction}

The coset construction [\cgko] of conformal field theories (CFTs)
[\cbpz,\cfms] has
proved to be a practical method for constructing
rational conformal field theories. Indeed,
 it may be that all rational
CFTs have a coset realisation.
It is of central importance to identify the
spectrum of primary fields present in such theories.
For an affine algebra $\hat g$ with subalgebra $\hat h$, the
primary fields of the coset CFT $\hat g/ \hat h$ can be labelled
$\phi^{\La}_\la$ where $\La$
and $\la$ are highest weights of the Lie algebras $g$ and $h$
respectively.
Not all pairs of labels $(\La,\la)$
give genuine and distinct primary fields but some combinations do
not correspond to fields present in the coset, and some
combinations of labels are equivalent [\cmoore].
In examining the spectrum of primary fields in a coset CFT, the
procedure, due to Schellekens and Yankielowicz [\cshellb],
of introducing an
``identification'' current, has proved extremely useful.
However, by examining the specific examples
$\widehat{su}(3)_2/\widehat{su}(2)_8$,
$(\hat E_6)_2/(\hat C_4)_2$,$(\hat E_7)_2/\asu(8)_2$ and
$(\hat E_8)_2/(\hat D_8)_2$ in detail
we show that this procedure is not always applicable, contrary to various
suppositions.
In these examples we find that extra identifications
and the vanishing of branching functions occur.
In addition we demonstrate that the series of cosets
$\asu(N)_2/\aso(N)_4$ and $\aso(2N)_2/\aso(N)_2\times\aso(N)_2$
also have non simple current identifications, and
exhibit additional null branching functions. We shall study the
series $\asu(N)_2/\aso(N)_4$ in detail.  The list of such ``Maverick''
cosets we have  found
has an A-D-E classification.

\vskip .4 truecm
\secta
\noindent
{\bf Review of the GKO construction}

Here we briefly review the Goddard, Kent and Olive (GKO) [\cgko]
construction for rational conformal field theories
and the Schellekens-Yankielowicz [\cshellb] mechanism for
obtaining the primary fields of coset models, through the use of
an identification current.

Consider the Kac-Moody algebra $\hat g$ [\ckac],
associated with the Lie algebra $g$.
Within the extended algebra of the Kac-Moody
algebra there is a Virasoro algebra,
for which the stress-energy tensor, $T_g$,
is formed using the
Sugawara construction [\cgod].
The central charge, $c_g$, is related to
the integer level, $k$, of the Kac-Moody algebra by
$$
c_g = { k \dim g  \over k+\tilde h }
\equn
$$
where $\tilde h$ is the dual Coxeter number of the Lie algebra $g$.

Suppose $g$ has a subalgebra $h$ so that correspondingly
$\hat g$ has
a subalgebra $\hat h$.
We can then form a new stress-energy tensor  [\cgko]
$
T_{g/h} = T_g - T_h
$
which satisfies the O.P.E. for an energy momentum
tensor with central charge
$
c_{g/h} = c_{g}- c_{h}
$.
This construction of the energy momentum tensor is the
well known GKO construction for coset conformal field theories.
Using this construction it is possible to construct
many CFTs with relatively small central charge, and it may be
that all rational CFTs have a coset realisation.
The irreducible
representations of $\hat g$
are labeled by the highest
weights of $g$, $\Lambda$,
At level $k$,
only those representations satisfying $\psi\cdot \Lambda \leq k$, where
$\psi$ is the highest root of $g$, are allowed.
The conformal weight of the
associated primary field is given by
$$
h_{\Lambda} =
{ \Lambda^2 + 2\La\cdot\rho_{g} \over
2( k+\tilde h)  }
\equn
$$
where $\rho_g$ is half the sum of the positive roots of $g$.

For a representation, $\Lambda$, of the Kac-Moody algebra the
restricted character is defined by
$$
\chi_\La= \tr_{\La} ( q^{L_0})
\; .
\equn
$$
Since $L_0^{g}=L_0^{h}+L^{g/h}_0$ it follows that
the characters of $g$ decompose into products of characters of
$h$ and $g/h$,
$$
\chi_{\Lambda} ( \tau )
=\sum_{\lambda}  \chi^{\Lambda}_{\lambda} ( \tau )
\; \chi_{\lambda} ( \tau  )
\; .
\equn\put\EQbeq
$$
The functions $\chi^{\Lambda}_{\lambda}(\tau)$ are the
``branching functions\quotesr\
of the coset CFT.
Not all pairs
of labels $(\La,\la)$ give rise to non-zero branching functions and not
all distinct labels give rise to distinct branching functions
[\cmoore,\cgep].
{}From \EQbeq\ the $h$-values of such a field is given by
$$
h_\lambda^\L=h_\L-h_\lambda+n
\put{\EQbeqn}\equn
$$
where $n$ can be calculated once \EQbeq\ has been solved for
$\chi^{\La}_{\la}$.

The Schellekens-Yankielowicz mechanism [\cshellb]
for deciding which
fields are non-zero and inequivalent is to use an \lq\lq identification
current\quotesr\
which is defined in terms of simple currents of the factors $\hat g$
and $\hat h$.
A simple current of a general CFT, $J$,
is a primary field with the simple fusion rules
[\cbpz,\cverlinde]
$
J \cdot \phi = \phi'
$.
For a rational CFT with a finite
number of primary fields there must be an integer
$N$ such that $J^N=1$.
In general the action of $J$ upon a primary field
$\phi$ will yield  fields $\{J^r \phi, r=0,1,\dots, N_\phi-1\}$,
where $J^{N_\phi}\phi=\phi$. The
integer $N_\phi$ must be a divisor of $N$.
When $J$ has integer conformal weight, that is $h(J)$ is integer.
(In general it can be shown that $h(J)=r/N$),
the CFT has a non-diagonal modular invariant (NDMI) [\cshella,\cintr]

whose form has been suggested to be the diagonal modular invariant
of an extended
algebra [\cmoore,\cgep,\cshella-\cextended].
Most Kac-Moody algebras contain simple currents,
some examples of which are
given in following sections.

The relationship between the characters of the coset algebra
and the branching functions has been the source of some confusion,
but has been elegantly resolved by Schellekens and Yankielowcz [\cshellb].
This relies on the observation that the diagonal combination of
characters,
$
Z= \sum_{a} \chi_a \chi_a^*
$
where the summation runs over all genuine chararacters of the coset CFT,
must be modular invariant [\ccardya].
Expresing $\chi_a$ as a sum
of the branching functions,
one should look for modular invariant combinations of the
branching functions.
To look for such modular invariants, note that the
branching functions of $\hat g/\hat h$ transform
as the characters of $\hat g\times \hat h^*$.
Hence if one can find a suitable modular invariant for
$\hat g \times \hat h^*$ then the corresponding object
for $\hat g / \hat h$ will be modular invariant and a candidate for
the diagonal modular invariant of the coset.
Schellekens and Yankielowicz generate such a
modular invarient using a simple current
$\phi^{J_1,J_2^*}$ of $\hat g \times \hat h^*$.
The corresponding field of
of $\hat g/\hat h$.
denoted $J_I$, is called the identification current.
After determining
$J_I=\phi^{J_1}_{J_2}$, the non-zero branching functions are those
which have
$$
h( J_I \cdot \phi^{\La}_{\la}  )
- h( \phi^{\La}_{\la}  ) = 0 \hskip 0.5 truecm \mod(1)
\equn
$$
and we have the following equivalence
$$
\phi^\La_\la \equiv
\phi^{J_1\cdot \La}_{J_2\cdot \la}
\equn
$$
The details of the identification current,
for a variety of cosets is given in [\cshellb].
The first condition is equivalent to requiring that $\la$
occurs within the representation $\La$ of $\hat g$.
This is often refered to as a conjugacy class relation as it is equivalent to
requiring $\La-\la'$ is a root of $g$. ( Where $\la'$ is the embedding $\la$.)
Obviously unless this is satisfied, the charachter $\chi^{\La}_{\la}$ is zero.
It is an explicit and clear assumption of the identification current method
that
branching functions only vanish when this relationship is not satisfied. (as we
shall see
additional branching functions do vanish.)

As a simple example, for $\asu(2)_k$ the simple current is $(k)$ which
satisfies $(k)\cdot(l)= (k-l)$. For $k$ odd there are no fixed points.
For $\asu(2)_k\times \asu(2)_1/ \asu(2)_{k+1}$ (a realisation of the
minimal models) the fields are $\phi^{l_1,l_2}_{l_3}$. The condition
for the branching function to be non-zero reduces to
$l_1+l_2-l_3 =0 \quad\mod(2)$, with the equivalence
$$
\phi^{l_1,l_2}_{l_3} \equiv \phi^{k-l_1,1-l_2}_{k+1-l_3}
\equn
$$
which can be
rearranged as the standard labelling for the fields of the
minimal models.

The identification current method is elegant and applies to
the majority of cosets,
however, as we shall show in the next section, there
exist cosets which
cannot be described purely in terms of such an identification current.


\vskip .4 truecm
\secta
\noindent
{\bf Maverick Cosets. }

In this section we present a class of coset theories
whose spectrum is not described by the Schellekens-Yankielowicz proceedure.
In general these theories have a smaller
spectrum of primary fields (h-values) than would be expected.
This arises because both more branching functions are non-zero than predicted
and more identifications occur. This suggests a larger symmetry than that
used
in the identification current method.

The simplest ``maverick'' coset is
$\asu(3)_2/\asu(2)_8$
[\cprevious]. As was observed in ref.~[\cprevious], where the characters of
coset theories was studied,
this model has more zero branching functions than predicted
and,  correspondingly, more equivalences.
Since $c=4/5$ for this model, the spectrum is entired specified
[\ccappelli] and the spectrum predicted by the
identification current method is
clearly not viable.
The identification current is given by
$$
J_I=\phi^{(00)}_8
\equn
$$
and its fusion rules are
$$J_I\cdot\phi^\La_\la=\phi^\La_{8-\la}
\equn
$$
Requiring $h^\La_\la-h^\La_{8-\la}$ to be an integer constrains
$\la$ to be even. This is the only selection rule resulting
from the simple current mechanism, or equivalently from
conjugacy class considerations. However [\cprevious] evaluation
of the characters
( see table~\Tabethree )
gives
extra identifications and nontrivial
vanishing of characters, that is to say nontrivial
selection rules.
After, taking the extra identifications into account the spectrum
matches perfectly that expected of the $c=4/5$ minimal model
in the non-diagonal modular invariant (NDMI) version.
This is also the first element of the
$W_3$ minimal series [\czam].
In the spirit of the identification current method we might
expect there to be a non-trivial NDMI for the theory $\asu(3)\times\asu(2)^*$
reflecting the extra
symmetry. Such a NDMI does exist, and is
$$ \eqalign{
Z= &\Bigl| \chi^{(00)}_0 +\chi^{(00)}_8 +\chi^{(11)}_4 \Bigr|^2
+\Bigl| \chi^{(11)}_2 +\chi^{(11)}_6 +\chi^{(00)}_4 \Bigr|^2
+\Bigl| \chi^{(10)}_2 +\chi^{(10)}_6 +\chi^{(02)}_4 \Bigr|^2  \cr
&+\Bigl| \chi^{(01)}_2 +\chi^{(01)}_6 +\chi^{(20)}_4 \Bigr|^2
+\Bigl| \chi^{(20)}_0 +\chi^{(20)}_8 +\chi^{(01)}_4 \Bigr|^2
+\Bigl| \chi^{(02)}_0 +\chi^{(02)}_8 +\chi^{(10)}_4 \Bigr|^2
\cr} \equn
$$
This NDMI only contains characters whose corresponding branching functions
are non-zero and also reflects the identifications,
for example $ \chi^{(00)}_0 \equiv \chi^{(00)}_8 \equiv \chi^{(11)}_4$
However this NDMI is
not generated by any simple current,
as can be seen, for example, by looking at the component
$\chi^{00}_0+\chi^{00}_8+\chi^{11}_4$. The last term is not related to
the others by any simple current, and indeed is a fixed point of
the identification current.
If we consider the branching function
$\chi^{11}_4$ which is equivalent to the identity,
then the h-value is
$h^{(11)}_4=h_{(11)}-h_4+n$ where $n$ is some integer.
However the branching rule in the Lie algebra
is
$$
(11)\rightarrow 2 \oplus 4
\equn
$$
so the $4$ of $su(2)$ occurs at the top level of the $(11)$
of $\asu(3)_2$, and thus $n$ is zero. This is confirmed by the
computation of the characters in table~\Tabethree .
We thus find that $h(\phi^{(11)}_4)=0$
and therefore this field must correspond to the identity.
Thus it is possible to recognise $\asu(3)_2/\asu(2)_8$ as having
a non trivial identification merely through examining
the finite dimensional Lie algebra branching rules
and looking for extra $h=0$ currents, as the
primary field corresponding to the identity in the
fusion algebra must be unique. We thus avoid having to make a
full evaluation of the branching functions. It is this
feature which is exploited to obtain new examples.

By examining lists of cosets such as found in [\cpatera],
and searching for examples of fields which must have, unexpectedly,
$h=0$
we have found the following examples
$$
\eqalign{
&\asu(N)_2/\aso(N)_4\cr
&\aso(2N)_2/\aso(N)_2\times\aso(N)_2\cr
&(\hat E_6 )_2/(\hat C_4)_2\cr
&(\hat E_7)_2/(\hat A_7)_2\cr
&(\hat E_8)_2/(\hat D_8)_2\cr
}
\equn
$$
In
all these examples have the Lie algebra branching rule for the adjoint
of $\hat g$, $\psi_g$,
$$
{\psi_g}={\psi_h}\oplus \la
\equn
$$
where there is only a single term $\la$ on the r.h.s.
Since this decomposition must occur at the top level in the representation
we can deduce
the integer shift is zero for the two branching functions
$\chi^{\psi_g}_{\psi_h}$ and
$\chi^{\psi_g}_{\la}$.
With this
information we can compute the h-value of the corresponding primary
fields.
For these examples we find
$h_{\psi_g}=h_\la$ implying $h(\phi^{\psi_g}_{\la})=0$
and
hence
$$
\phi^{\psi_g}_\la\equiv\phi^{(00)}_0
\equn
$$
The three exceptional cases have central charge less
than one so we can immediately check whether the identification current
mechanism
gives us a permissable set of weights.

For the $E_7/A_7$ case $c=7/10$ and the
spectrum of $h$-values of the fields in this minimal
model is
$\{ 0,{3\over80},
{1\over10},{1\over2},{7\over16},{3\over5}\}$ [\ccappelli].
Applying the simple current mechanism with
$$
J_I=\phi^{2\la_6}_{2\la_2}
\equn
$$
we obtain the $h$-values, up to integer shifts, in table~\Tabeseven .
The $h$-values indicated with a $\dagger$  are not permissable
and  must correspond to vanishing branching functions.
Notice, that extra identification clearly exist for all fields.
This model therefore is clearly a maverick with a considerably
smaller spectrum ( 6 fields ) than expected ( 18 fields ).
If one were using this model, for example,
for superstring model building one would clearly be
led to wrong conclusions on the spectrum using the standard identification
current method.

If we examine the case $(\hat E_6)_2/(\hat C_4)_2$,
we have $c=6/7$.
There appear to be numerous
illegal $h$-values, as can be seen in table~\Tabesix .
Those  $h$-values in table~\Tabesix\ which are permitted are
 $\{0,{1\over21},
{2\over7},{1\over3},{10\over21},{6\over7}\}$.
This list corresponds exactly to that for the NDMI of the $c=6/7$
minimal model. Hence this example is also a maverick theory.
The final exceptional case $(\hat E_8 )_2 / (\hat D_8 )_2$
has $c=1/2$ and corresponds
to the first minimal model.  The case of $E_8$ is slightly trickier to
deal with. In general, the identification current is valid for all
values of the
level $k$. However, $E_8$ has an exceptional simple current which only
exists for $k=2$ [\cshellc].
One could envisage incorporating this current into the identification
current purely for $k=2$. This is difficult to make work, however, because the
extra simple current has half-integer weight and
one is led to the conclusion that
this model is also a maverick.

Let us now consider the sequence of theories $\suson$.
Here the embedding is specified by the branching
rule
$$
\psi\rightarrow\psi\oplus s
\equn
$$
where $s$ denotes the spinor representation of $so(N)$.
We thus find  $\phi^\psi_s\equiv\phi^0_0$.
Similarly in $\dser$ we find $\phi^\psi_{s,s}\equiv\phi^0_0$.
Since both these series have $c={2(N-1)\over N+2}$ and $c=1$
respectively we cannot immediately
see which $h$-values, and therefore fields, are allowed.
In order to do so  it would be necessary to construct
the modular S-matrix explicitly on a case by case basis.
Alternatively one could attempt to identify the model involved in
general.
If we examine the spectrum of h-values in the cosets
$\suson$ for $N$ even then we find a large number of field
which are equivalent to the identity field.
Specifically
$$
\chi^{(1,0,0,\cdots 0,0,1)}_{(2,0,0,\cdots, 0,0,0)}
\equiv
\chi^{(0,1,0,\cdots 0,1,0)}_{(0,2,0,\cdots, 0,0,0)}
\equiv
\chi^{(0,0,1,\cdots 1,0,0)}_{(0,0,2,\cdots, 0,0,0)}
\equiv
\chi^{(0,\cdots 1,0,1,\cdots)}_{(0,0,,\cdots, 0,2,2)}
\equiv
\chi^{(0,\cdots 0,2,0,\cdots)}_{(0,0,,\cdots, 0,0,4,)}
\equiv
\chi^{(0,\cdots 0,2,0,\cdots)}_{(0,0,,\cdots, 0,4,0,)}
\equn
$$
This list includes those identified with the identity by the
identification current proceedure.
This large class of fields which are equivalent to the
identity is indicative of a large symmetry between the characters.
For all cases when $\phi^{\La}_{\la}$ is equivalent to $\phi^0_0$
then $\La$ is a member of the root lattice. In fact if we look at the set of
coincidence we find that whenever $\phi^{\La}_{\la}\equiv\phi^{\La'}_{\la'}$
then we have
$\La-\La'$ in the root lattice. If we look at the list of fields above then we
find that
the set of $\{ \La \}$ at level $2$ differing from $0$ is in fact saturated.
Before looking at the general structure of the $\suson$ series we will look at
the
cases $\asu(4)_2/\asu(2)_4\times\asu(2)_4$ and $\asu(5)_2/\aso(5)_4$. The first
case
is a rather special ``low $N$'' case which we shall examine in detail and the
second shall
give us the flavour of the series.
For the case $\asu(4)_2/\asu(2)_4\times\asu(2)_4$ the embedding of the
$\asu(2)_{2k}\times\asu(2)_{2k}$
within $\asu(4)_k$ is given by
$$
\eqalign{
J^{\pm}_m &= ( J_m^{\pm \alpha_1} + J_m^{\pm\alpha_3} )
\; ; \hskip 1.8 truecm
H_m = \biggl( H_m^1 - \sqrt{ 1 \over  3} H_m^2 +\sqrt{ 2 \over3} H_m^3 \biggr)
\cr
\bar{J}{}^{\pm}_m &= ( J_m^{\pm (\alpha_1+\alpha_2)} +
J_m^{\pm(\alpha_2+\alpha_3)} )
\; ; \bar{H}_m = \biggl( \sqrt{ 4 \over  3} H_m^2 +\sqrt{ 2 \over3} H_m^3
\biggr)
\cr}
\equn
$$
With this embedding one may calculate the branching functions for the cosets as
given in
ref.~[\cprevious]. One finds that there are zero characters and extra
equivalences beyond
that expected from the identification current method.
For this coset there is a pair of identification curents
$$
J_1 = \phi^{(020)}_{4,0} , \hskip 1.5 truecm
J_2 = \phi^{(020)}_{0,4}
\equn
$$
In table~\Tabefour\ the spectrum predicted
using these identification currents is shown.
Also, from a direct analysis of the branching functions
[\cprevious] we have zero characters as indicated in the table.
In addition there are correspondingly extra equivalences.
{}From
the branching functions alone there is a little ambiguity in
regards the conjugate
weights within $su(4)$. This we have resolved so that $\La-\la$ is a root. This
is consistent
with the $\asu(3)_2/\asu(2)_8$ case where this choice was vindicated by the
non-diagonal modular invariant of $\asu(3)_2 \times \asu(2)_8^*$.

One notable feature of the list is the branching function $\chi^{(101)}_{2,2}$.
This branching
function
may be found by inspection to
be the sum of two separate characters of the coset. The branching function
must contain the identity since $h=0$.
We find that
$$
\eqalign{
\chi_{2,2}^{(101)}
&= 1 + q +2q^2 +5q^3 +7q^4 +10q^5 \cdots  \cr
&=( 1 + q^2 +2q^3 +4q^4 +5q^5  \cdots )
 +q( 1 +q +3q^2 +3 q^3 +5q^4 + \cdots ) \cr
&=\chi^{(000)}_{0,0} + q \chi^{(000)}_{0,4}
\cr}
\equn
$$
This is a rather
surprising feature. It is usual that branching functions
$\chi_{\la}^{\La}$
correspond to irreducible characters of the coset but this provides
a counter-example
to this intuition.
This feature will relate to the existance of
fixed points of the identification current [\cshellc].

We now turn to a simpler case,
$\asu(5)_2/\aso(5)_4$. The spectrum of $h$-values
determined using the  identification
current method is given in table~\Tabefive.
In this case we will not calculate the
branching function explicitly but shall `deduce`
or postulate the spectrum. Since we know
$\phi^{(1001)}_{(20)}$ is equivalent to the identity
definately, plus various other
equivalences we postulate that
only the fields indicated are non-zero and basically
each field in the coset has three different labelings
(in addition to those for
the identification current.) For example,
$\phi^{(0000)}_{(00)}\equiv\phi^{(1001)}_{(02)}\equiv\phi^{(0110)}_{(20)}$
and
$\phi^{(0200)}_{(20)}\equiv\phi^{(0001)}_{(12)}\equiv\phi^{(1010)}_{(02)}$.
In the second case we had a choice of identifications due to the conjugacy (
i.e. to take $(1010)$ or $(0101)$). We have been able to do so by requiring
that $\La-\La'$ is a root in all cases.  For this example we could have
identified the spectrum of this coset by selecting the fields determined by the
identification current and then ordering the set of $\{ \La\}$
into conjugacy classes
under the root lattice.
The entire spectrum could then be determined by only using a single
$\{ \La\}$ from each conjugacy class.
With this proceedure we obtain a spectrum of $h$-values
$(0,(2/35)^2,(3/35)^2,(1/5)^2,(2/7)^2,(17/35)^2, (23/35)^2, (4/5)^2, (6/7)^2,)$
where the $h$-values are quoted up to an integer part.
Have we any confidence that this spectrum is genuine?. In fact this spectrum
matches prescisely that of the
first element of the minimal $W_5$ extended algebra.
In general the
$\asu(N)_2/\aso(N)_4$ has the same $c$-value and,
after making the postulated changes to the
proceedure for obtaining the spectrum,
spectrum of $h$-values as the first element of
the $W_N$ minimal series,
$(\asu(N)_1\times\asu(N)_1)/\asu(N)_2$.  When one looks at the examples
$\asu(3)_2/\asu(2)_8$ and $\asu(4)_2/\asu(2)_4\times\asu(2)_4$
where we have calculated the branching
functions we find the characters
match exactly those for the first elements of the
minimal series of $W_3$ and $W_4$ algebras respectively.

The examples we have found have a very simple A-D-E classification
These cosets are $\hat g_2/ \hat h_{2k'}$ , one for each $\hat g$ in the A-D-E
series.
These models are postulated to be
equivalent to the first elements of the minimal series
$\hat g_1 \times \hat g_1 / \hat g_2$.

Although a proof is lacking,
we have studied carefully models outside this classification.
For example, we have take $\hat G_2 / \hat h$ for all possible maximal $\hat h$
( $\asu(3)$, $\asu(2)$, $\asu(2)\times\asu(2)$)
and have found the spectrum generated by
the identification current methods to be correct.
We have no examples where $\hat g$ is other
than simply laced and at level 2.

\vskip 0.4cm
\noindent
\secta
{\bf  Conclusions}

In examining the spectrum of
primary fields of coset theories,
we have found that coset conformal theories have a more subtle
structure than previously understood
and have found a class of cosets with a ``maverick'' spectrum.
In these theories, there is a smaller spectrum of primary fields
than predicted by the identification current method indicating a larger
symmetry than used in that method. The maverick behaviour only occurs
for  the level equal to two ; for higher levels the spectrum being
decribed correctly by the identification current method.
The smaller spectrum arised firstly, from the vanishing of branching
functions other than that expected purely from conjugacy class selection
rules and, secondly, from the extra equivalences amongst fields.

\noindent
The examples given all have the following list of properties in
common:

1) $G/H$ is a symmetric space.

2) $ \hat h_{k'} \subset \hat g_1$ is a conformal embedding.

3) $g$ is simply laced.

4) $\hat g$ is at level  $k=2$.

5) An additional $h=0$ field $\phi^\psi_\la$ is uniquely determined
by the Lie algebra branching rule specifying the embedding.

6) $\phi^\psi_\la$ is a fixed point of the identification current.

\vskip 0.2 truecm
Exactly how these properties are related to the existence
of a sufficient number of null states in the representations
of $\hat g$ that we have additional null baranching functions is
not apparent. It would be significant if one could find an example which
does not satisfy the above criteria.
The extra symmetry responsible for maverick behaviour
is perhaps related to the $W_N$ and related algebras.

The majority of coset models have, of course,
a spectrum determined by the identification
current.
The models we have found ( by an exhausting if not exhaustive search
of cosets in [\cpatera] ) have
a classification in terms of the A-D-E series, which appears in so many diverse
areas of physics.
For each element, $g$, of the A-D-E series we have precisely one example
$\hat g / \hat h$ which is a maverick.
This classification is very intreging and may lead to an
understanding of the occurance of Maverick coset and
eventually a better understanding
of coset theories in general.

This work was supported by a S.E.R.C advanced fellowship
a NATO grant CRG-910285 (D.C.D)
and a S.E.R.C studentship (K.G.J). We thank
John Gracey and David Olive
for helpful discussions.

\vfill\eject

\vskip 0.4cm
\noindent
\secta
{\bf Tables}

\vskip -0.5 truecm
$$
\vbox{\offinterlineskip
\hrule
\halign{ &\vrule# & \strut\quad\hfil#\quad\cr
height2pt&\omit&&\omit&&\omit&&\omit&\cr
& Equivalences of Character   &&Character Value\qquad\qquad&\cr
height2pt&\omit&&\omit&&\omit&&\omit&\cr
\noalign{\hrule}
height2pt&\omit&&\omit&&\omit&&\omit&\cr
&\quad$\chi_{0}^{(00)} \equiv \chi_{8}^{(00)}\equiv \chi_{4}^{(11)}\hfill$
&& $1 ~~~~~~ +q^2 +2q^3 +3q^4 +4q^5 \cdots  $   &\cr
&\quad  $\chi_{2}^{(11)} \equiv \chi_{6}^{(11)}\equiv \chi_{4}^{(00)}\hfill$
&& $ 1+2q +2q^2 +4q^3 +5q^4 +8q^5 \cdots $   &\cr
&\quad $\chi_{2}^{(10)} \equiv \chi_{6}^{(10)}\equiv \chi_{4}^{(02)}\hfill$
&& $1 +q +2q^2 +3q^3+5q^4 +7q^5 \cdots $   &\cr
&\quad $\chi_{2}^{(01)} \equiv \chi_{6}^{(01)}\equiv \chi_{4}^{(20)}\hfill$
&& $1 +q +2q^2 +3q^3+5q^4 +7q^5 \cdots $    &\cr
&\quad $\chi_{0}^{(20)} \equiv \chi_{8}^{(20)}\equiv \chi_{4}^{(01)}\hfill$
&&
$1 +q +2q^2 +2q^3 +4q^4 +5q^5 \cdots $
&\cr
&
$\quad \chi_{0}^{(02)} \equiv \chi_{8}^{(02)}\equiv \chi_{4}^{(10)}\hfill$
&&
$1 +q +2q^2 +2q^3 +4q^4 +5q^5 \cdots $
&\cr
&\quad $\chi_{2}^{(00)} \equiv \chi_{6}^{(00)}\hfill $
&&$~$ 0 \hfill  &\cr
&\quad  $\chi_{2}^{(20)} \equiv \chi_{6}^{(20)}\hfill $
&&$~$ 0 \hfill  &\cr
&\quad  $\chi_{2}^{(02)} \equiv \chi_{6}^{(02)}\hfill $
&&$~$ 0 \hfill  &\cr
&\quad  $\chi_{0}^{(10)} \equiv \chi_{8}^{(10)}\hfill $
&&$~$ 0 \hfill  &\cr
&\quad $\chi_{0}^{(01)} \equiv \chi_{8}^{(01)}\hfill $
&&$~$ 0 \hfill  &\cr
&\quad  $\chi_{0}^{(11)} \equiv \chi_{8}^{(11)} \hfill$
&&$~$ 0 \hfill  &\cr
height2pt&\omit&&\omit&&\omit&&\omit&\cr}
\hrule}
$$
{\bf Table~\Tabethree.} For the coset theory
${\widehat{su}(3) }_2/{\widehat{su}(2)}_8$,
we show the extra equivalences and vanishing of characters beyond
that expected by the identification current method.
Using the identification currrent method all characters shown
are expected
to be non-zero and only the first equivalences are expected.

\vskip 1.3 cm


\def\la{\lambda}
\vskip 0.4cm
\def\hozline{height2pt&\omit&&\omit&\omit&\omit&&\cr
\noalign{\hrule}
height2pt&\omit&&\omit&\omit&\omit&&\cr}

\vbox{\offinterlineskip
\hrule
\halign{\vrule #&\quad\hfil$#$\hfil\quad&
\vrule#&\quad\hfil$#$\hfil\quad&
\quad\hfil$#$\hfil\quad&\quad\hfil$#$\hfil\quad
&\quad\hfil$#$\hfil\quad&#\vrule\cr
height2pt&\omit&&\omit&\omit&\omit&&\cr
&\omit&&\multispan4\hfil$ E_7$ weight label for orbit\hfil&\cr
\hozline
& A_7\hbox{\rm\  weight}
&&0&\la_1&\la_6&\la_7&\cr
\hozline
&0&&0 &{ 9 \over10}^\dagger  &\omit&\omit&\cr
\hozline
&2\la_2&&{1\over2} &{2\over5}^\dagger &\omit&\omit&\cr
\hozline
&\la_4&&{1\over10}  & 0 &\omit&\omit&\cr
\hozline
&\la_2+\la_6&&{3\over5} & {1\over2} &\omit&\omit&\cr
\hozline
&\la_1+\la_3&& {7\over10}^\dagger  &{3\over5} &\omit&\omit&\cr
\hozline
&\la_3+\la_5&&{ 3\over5} & {1\over2} &\omit&\omit&\cr
\hozline
&\la_2&&\omit&\omit& {3\over80} & { 51\over80}^\dagger &\cr
\hozline
&\la_1+\la_5 &&\omit&\omit& {7\over16} &
{3\over80}&\cr
\hozline
&2\la_1&&\omit&\omit&{67\over80}^\dagger &{7\over16}&\cr
height2pt&\omit&&\omit&\omit&\omit&&\cr}
\hrule}
\vskip 0.4cm
\noindent
{\bf Table~\Tabeseven .} The $h$-values for the coset
$(\hat E_7)_2/(\hat A_7)_2$ after calculating the spectrum using
the identification current.
Only the non-zero inequivalent fields are
shown.
Those
shown with a $\dagger$ are {\it not} part of the spectrum
of the $c=7/10$ minimal model and hence must vanish.
\vskip 0.4cm


\def\la{\lambda}
\def\hozline{height2pt&\omit&&\omit&\omit&\omit&\omit&\omit&\omit&\cr
\noalign{\hrule}
height2pt&\omit&&\omit&\omit&\omit&\omit&\omit&\omit&\cr}

\vbox{\offinterlineskip
\hrule
\halign{\vrule #&\quad\hfil $#$ \hfil\quad
&\vrule#&\quad\hfil $#$ \hfil\quad&
\quad\hfil $#$ \hfil\quad&
\quad\hfil $#$ \hfil\quad&\quad\hfil $#$ \hfil\quad&
\quad\hfil $#$ \hfil\quad&\quad\hfil $#$ \hfil\quad&#\vrule\cr
height2pt&\omit&&\omit&\omit&\omit&\omit&\omit&\omit&\cr
&\omit&&\multispan6\hfil$ E_6$ weight label\hfil&\cr
\hozline
&\omit&&0&\la_1\sim\la_5&\la_6&\la_2\sim\la_4&2\la_1\sim2\la_5
	&\la_1+\la_5&\cr
\hozline
& C_4\hbox{\rm\  weight} && & & &
& & &\cr
\hozline
&0 &&0*&{13\over21}&{6\over7}*&{25\over21}&{4\over3}*& {2\over7}* &\cr
\hozline
&\la_4 &&{1\over7}&{16\over21}&{0}*&{1\over3}*&{10\over21}*
&3\over7&\cr
\hozline
&\la_1+\la_3 &&{6\over 7 }*&{10\over21}* &5 \over7&{1\over21}*
&4\over 21&1\over 7&\cr
\hozline
&2\la_1 &&{2\over7}*&{19\over21}&1\over7&{ 10\over21}*
&13\over21&4\over7&\cr
\hozline
&2\la_2 &&{5\over7}&{1\over3}*&4\over7&19\over21
&{1\over21}* &0*&\cr
height2pt&\omit&&\omit&\omit&\omit&\omit&\omit&\omit&\cr}
\hrule}
\noindent
{\bf Table~\Tabesix .} The $h$-values (up to integer
shift) for the coset $(\hat E_6)_2/(\hat C_4)_2$.
Those field indicated $*$ appear in the $c=6/7$ minimal model.
All other branching functions must vanish.
\vskip 1.4cm

\vskip 1.5 truecm


\def\la{\lambda}
\def\hozline{height2pt&\omit&&\omit&\omit&\omit&\omit&\omit&\omit&\cr
\noalign{\hrule}
height2pt&\omit&&\omit&\omit&\omit&\omit&\omit&\omit&\cr}

\vbox{\offinterlineskip
\hrule
\halign{\vrule #&\quad\hfil $#$ \hfil\quad
&\vrule#&\quad\hfil $#$ \hfil\quad&
\quad\hfil $#$ \hfil\quad&
\quad\hfil $#$ \hfil\quad&\quad\hfil $#$ \hfil\quad&
\quad\hfil $#$ \hfil\quad&\quad\hfil $#$ \hfil\quad&#\vrule\cr
height2pt&\omit&&\omit&\omit&\omit&\omit&\omit&\omit&\cr
&\omit&&\multispan6\hfil$ SU(4)$ weight label\hfil&\cr
\hozline
&\omit&& (000)    & (100) & (001) & (200)
	& (101)  & (010) &\cr
\hozline
& SU(2)\times SU(2) \hbox{\rm\  weight}
&& & & & & & & \cr
\hozline
& 0 ,0  && 0 &\cdot & \cdot &{3\over4 }&{8\over12 }^{\dagger}
&{5\over12}^{\dagger} &\cr
\hozline
& 0, 2 &&{2\over3}^{\dagger}& \cdot & \cdot
&{5\over12}^{\dagger} &1\over3
&1\over12 &\cr
\hozline
& 0 , 4 &&{0 }& \cdot  & \cdot  & 3 \over 4
&{8 \over 12}^{\dagger} &{5\over 12}^{\dagger} &\cr
\hozline
& 1 , 1   &&\cdot &{1\over 16 }&1\over 16 & \cdot
& \cdot  & \cdot  &\cr
\hozline
& 1 , 3 && \cdot &{9\over 16}& 9\over 16 & \cdot
& \cdot &  \cdot  &\cr
\hozline
& 2 , 0  &&{ 2\over3}^{\dagger}  &\cdot & \cdot
& {5 \over 12}^{\dagger}
&1\over 3 & { 1 \over 12 }&\cr
\hozline
&  2 , 2  &&{1 \over 3} &  \cdot & \cdot  &1\over 12
& 0 & { 3 \over 4 } & \cr
height2pt&\omit&&\omit&\omit&\omit&\omit&\omit&\omit&\cr}
\hrule}
\vskip 0.4cm
\noindent
{\bf Table~\Tabefour .} The $h$-values for the coset
$\asu(4)_2/(\asu(2)_4 \times \asu(2)_4 )$.
The values shown correspond to the spectrum predicted by
the identification current method. Those values indicated by
$\dagger$ are in fact zero when explicitly evaluated.
Only one of each conjugate pair of $\asu(4)$ weights is shown.

\vskip 0.4cm

\vfill\eject

\def\quad{\hskip 0.7em\relax}

\def\la{\lambda}
\def\hozline{height2pt&\omit&&\omit&\omit&\omit
&\omit&\omit&\omit&\omit&\omit&\omit&\cr
\noalign{\hrule}
height2pt&\omit&&\omit&\omit&\omit
&\omit&\omit&\omit&\omit&\omit&\omit&\cr}

\vbox{\offinterlineskip
\hrule
\halign{\vrule #&\hfil $#$ \hfil&
\vrule#&
\hfil $#$ \hfil&
\quad\hfil $#$ \hfil&
\quad\hfil $#$ \hfil&
\quad\hfil $#$ \hfil&
\quad\hfil $#$ \hfil&
\quad\hfil $#$ \hfil&
\quad\hfil $#$ \hfil&
\quad\hfil $#$ \hfil&
\hfil $#$ \hfil&
#\vrule\cr
height2pt&\omit&&\omit&\omit&\omit&\omit
&\omit&\omit&\omit&\omit&\omit&\cr
&\omit&&\multispan6\hfil$ SU(5)$ weight label\hfil& &&& \cr
\hozline
&\omit&& (0000)    & (1001) & (0110) & (2000)
	& (1100) & (0100) & (0200) & (1000) & (1010)    & \cr
\hozline
& SO(5)  \hbox{\rm\  weight}&& & & &
& & &    & & & \cr
\hozline
& (0,0)   && 0\; * &{5\over 7}&{1 \over 7 }&{ 4 \over 5 }\; *
&{33 \over 35 } &{18 \over 35 } & { 1 \over 5}\; *  & {12\over 35}
&{32\over 35 } & \cr
\hozline
& (1,0)  &&{5\over 7}&{3\over7}&{6\over 7 } \; *
&{ 18 \over 35 }
&{ 23 \over 35} \; *
&{ 8 \over 35 }
& {32 \over 35 } & { 2\over 35 } \; *   & { 22 \over 35}& \cr
\hozline
& (2,0)  &&{2  \over 7 } \; * &{ 0 } \; * &{ 3\over7}
& {3  \over 35} \; *
&{ 8 \over 35 }
&{ 28\over  35}\; * & { 17 \over 35 } \; *  & {22\over 35}
&{  1 \over 5} \; *  & \cr
\hozline
& (0,2)   &&{4 \over7 }&{2\over7}\; * &{ 5 \over 7}&{ 13\over35}
& { 18 \over 35 }
& { 3\over 35 }\; *  &{ 27 \over 35} &{32\over 35}
& {17 \over 35 }\; *  & \cr
\hozline
& (1,2)  &&{1 \over7}&{6 \over7}\; * &{2 \over 7}\; *
&{ 33 \over 35}
&{ 3 \over 35 }\; *
&{ 23 \over 35}\; * &{ 12 \over 35 }
& {17\over 35 }\; *
&{2 \over 35} & \cr
\hozline
& (0,4)  &&{6 \over7}\; * &{ 4\over7}&{ 0 }\; *
&{23 \over 35} \; *
& { 28 \over 35 } \; *
&{ 13 \over  35} & {2 \over 35 }\; *
 & {1\over 5}\; *
&{27 \over 35 } & \cr
height2pt&\omit&&\omit&\omit&\omit&\omit&\omit&\omit&\omit&\omit&&\cr}
\hrule}
\vskip 0.4cm
\noindent
{\bf Table~\Tabefive .} The $h$-values for the coset
$\asu(5)_2 / \aso(5)_4 $. All values shown are allowed by the
identification current procedure. Those indicated
by $*$ are those postulated to correspond to genuine
non-zero branching functions.

\vskip 0.4cm

\vfill\break
\noindent

{\bf References }

\parindent=-13 pt
\parskip=0 pt

[\cgko] P.\ Goddard, A.\ Kent, D.\ Olive., Phys. Lett.
 {\bf 152} (1985) 88;
 Commun. Math. Phys. {\bf 103} (1986) 105.

[\cbpz]  A.A.\ Belavin, A.M.\ Polyakov and A.B.
Zamolodchikov,  Nucl.\ Phys.\ {\bf B241}(1984) 333;
J.\ L.\ Cardy, Nucl. Phys. {\bf B240} (1984) 514.

[\cfms]
D.\ Friedan, Z.\ Qiu and S.\ Shenker,
Phys. Rev. Lett. {\bf 52} (1984) 1575.

[\cmoore]
G. Moore and N. Sieberg, Phys. \ Lett.\ {\bf 220B} (1989) 422;
W.\ Lerche, C.\ Vafa and N.\ Warner, Nucl.\ Phys.\ {\bf B324} (1989) 673.

[\cgep] D.\ Gepner, Phys.\ Lett.\ {\bf 222B} (1989) 207.

[\cshellb]
A.N.\ Schellekens and S.\ Yankielowicz,
Nucl. Phys. {\bf B334} (1990) 67.

[\ckac]
V.G.\ Kac, Func.\ Anal.\ App.
{\bf 1} (1967) 328;
R.V.\ Moody, Bull.\ Am.\ Math.\ Soc.\
{\bf 73} (1967) 217;
K.\ Bardakci and M.B.\ Halpern, Phys.\ Rev.\
{\bf D3} (1971) 2493.

[\cgod] P. Goddard and D. Olive, Int. J. Mod. Phys. {\bf A1} (1986) 303.

[\cverlinde]
E.\ Verlinde, Nucl.\ Phys.\ {\bf B300} (1988) 360.

[\cshella]
A.N.\ Schellekens and S.\ Yankielowicz,
Nucl. Phys. {\bf B327} (1989) 673;
Phys.\ Lett. {\bf 227B} (1989) 387.

[\cintr]
A.K.\  Intriligator, Nucl. Phys. {\bf B332} (1990) 541.

[\cbais]
F. Bais, P. Bouwknegt, K. Schoutens and M. Surridge,
Nucl. Phys. {\bf B304} (1988) 348.

[\cextended] Z.\ Bern and D.C.\ Dunbar, Phys. Lett. {\bf 248B} (1990) 317.

[\ccardya]
J. Cardy, Nucl.\ Phys.\ {\bf B270}  (1986) 186.

[\cshellc]
A.N.\ Schellekens and S.\ Yankielowicz,
Int.J.Mod.Phys.{\bf A5} (1990) 2903:
Nucl. Phys. {\bf B366} (1991) 27.

[\cprevious] D.C.\ Dunbar and K.G.\ Joshi, to be published in
		Int. J. Mod. Phys.

[\ccappelli]
A.~Cappelli, C.~Itzykson and J.-B.~Zuber, Nucl.~Phys.\ {\bf B280}
(1987) 445.

[\czam] A.B. Zamolodchikov, Theor. Math. Phys. {\bf 65} (1986) 1205.

[\cpatera] R. Slansky, Phys.\ Rep. {\bf 79} (1981) 1 ;
W.\ Mackay and J.\ Patera, {\it Tables of Dimensions, Indices and
branching rules for simple Lie algebras}, (Dekker , New York 1981).

\vfill\eject
\bye